\documentclass[journal=jacsat,manuscript=article]{achemso}
\usepackage[breaklinks=true,colorlinks,citecolor=blue,linkcolor=blue,urlcolor=blue]{hyperref}
\usepackage[table,xcdraw]{xcolor}
\usepackage{epsfig, mathrsfs, color, latexsym, subfigure, marginnote, gensymb, amsmath}
\usepackage{graphicx}
\usepackage{xcolor}
\usepackage{booktabs}
\usepackage{adjustbox}
\usepackage{doi}

\author{Kammampati Sai Kumar}
\affiliation{Department of Materials Science \& Engineering, Indian Institute of Technology, Kanpur, Kanpur 208016, India}

\author{Albert Linda}
\email{albert20@iitk.ac.in}
\affiliation{Department of Materials Science \& Engineering, Indian Institute of Technology, Kanpur, Kanpur 208016, India}

\author{Shubham Kumar Maurya}
\affiliation{Department of Materials Science \& Engineering, Indian Institute of Technology, Kanpur, Kanpur 208016, India}

\author{Somnath Bhowmick}
\email{bsomnath@iitk.ac.in}
\affiliation{Department of Materials Science \& Engineering, Indian Institute of Technology, Kanpur, Kanpur 208016, India}
\date{\today}

\title{Physics Aware Representation Learning on Electronic Charge Density for Materials Property Prediction}

\begin{document}


\begin{abstract}
The fundamental quantity governing the mechanical and thermodynamic properties of a crystalline solid is its electronic charge density. Yet, its direct use for the rapid prediction of materials properties remains challenging due to its high dimensionality. Here, we present a physics-informed deep learning framework that directly predicts mechanical and thermodynamic properties from the three-dimensional electronic charge density derived from density functional theory (DFT). The proposed approach first utilizes a three-dimensional convolutional autoencoder for unsupervised dimensionality reduction, compressing a high-resolution charge-density grid ($128 \times 128 \times 128$) into a compact latent representation ($16 \times 16 \times 16 \times 16$) while preserving physically meaningful features, as confirmed by negligible reconstruction errors across diverse crystal systems. The compressed latent-space representation of charge density is then used by two different regression models for property prediction: Light Gradient Boosting Machine (LightGBM) and Attention-based 3D Convolutional Neural Networks (Att CNN), and their performance is compared. Combining composition-based descriptors (Material Agnostic Platform for Informatics and Exploration or MAGPIE) with electronic charge density data further improves the model accuracy. Using a dataset of about 6059 inorganic compounds spanning multiple crystal symmetries, the models achieve strong predictive performance for bulk modulus K ($R^2 = 0.94$), Young’s modulus E ($R^2 = 0.88$), shear modulus G ($R^2 = 0.87$), formation energy E$_{form}$ ($R^2 = 0.96$), and Debye temperature $\Theta$ ($R^2 = 0.89$). This work establishes electronic charge density as a transferable, physics-grounded descriptor for materials property prediction, requiring $\approx 1/25$ the computational resources of full-fledged DFT calculations.
\end{abstract}


\section{Introduction}
Machine learning (ML) has found widespread applications across materials science~\cite{Kumar2024, Sinha2025,alex2020,cuevas2020}, encompassing tasks such as crystal structure classification~\cite{ziletti2018insightful}, microstructure segmentation~\cite{deCost2020microstructure}, image denoising~\cite{ahmad2023denoising}, modeling of microstructural evolution~\cite{ahmad2023autoencoder, Ahmad_2025, Gaikwad2025}, and the generation of hypothetical materials with targeted properties~\cite{Dan2020}. Concurrently, density functional theory (DFT) has been pivotal in facilitating first-principles evaluations of fundamental material properties, such as elastic constants~\cite{Hafner2008, Linda_2023}, thereby accelerating materials design by efficiently screening candidate materials and minimizing experimental effort. Combining ML with DFT opens new avenues for designing materials with customized properties~\cite{sidnov2024machine, KIM2019124}.

ML has revolutionized computational materials science by providing robust frameworks for property prediction and the accelerated discovery of new materials. Among various approaches, ensemble methods are very popular due to their higher accuracy, reduced overfitting, and increased robustness to noise and outliers. Their effectiveness has been demonstrated for predicting elastic properties of BCC Ti and Zr alloys~\cite{sidnov2024machine} and elastic moduli of the high-entropy alloy Al$_{0.3}$CoCrFeNi~\cite{KIM2019124}. Another popular class is convolutional neural networks (CNNs), especially their three-dimensional (3D) variants, which have demonstrated remarkable ability to capture intricate correlations between a material’s electronic charge density and its macroscopic properties, such as elastic properties. Applications to FCC high-entropy alloys have demonstrated that these models can extract key information from the spatial arrangement of charge density that correlates strongly with elastic constants, providing a data-driven alternative to computationally intensive DFT calculations~\citep{mirzaee2024elastic}. Features learned from the spatial distribution of electron density are more transferable and robust to compositional changes than composition-based descriptors, which highlights the ability of charge density to capture intrinsic, physically meaningful structural patterns~\citep{Zhao2020}. 

Beyond direct property prediction, deep learning approaches utilizing three-dimensional (3D) data have found diverse applications across materials science. For instance, models trained on sparse 3D electron density distributions have been employed for symmetry classification in inorganic materials~\citep{Kim2024}. In another work, deep neural networks have been developed to predict charge densities~\citep{Kamal2020}. Recent studies have also explored optimized network architectures, such as lean convolutional networks, for mapping electron charge density to material properties, achieving reliable predictions with reduced computational cost~\citep{Ray2025}. Machine learning has further been applied to develop lossless multiscale constitutive models and to predict full elasticity tensors, demonstrating its potential to capture complex material responses across multiple scales~\citep{Mianroodi2022}. Additionally, high-throughput frameworks combining \textit{ab initio} calculations with machine learning have been introduced to predict the properties of hard-coating alloys~\citep{Levamaki2022}.

Working with high-dimensional data, such as the 3D charge densities of materials, poses a significant computational challenge. Autoencoders offer an elegant solution for dimensionality reduction, adept at learning compact, low-dimensional representations from such inputs in an unsupervised manner. The effectiveness of autoencoders for dimensionality reduction has recently been demonstrated, particularly in compressing sparse, high-dimensional scientific data~\cite{9680154, 9096747}. The resulting compressed latent space then serves as a highly efficient and informative input for downstream modeling, such as ensemble-based regression or 3D-CNN regression.  

Building upon this extensive body of work, this study introduces a novel, two-stage hybrid approach to predict material properties, including bulk modulus ($K$), shear modulus ($G$), Young's modulus ($E$), formation energy per atom ($E_{\text{form}}$), and Debye temperature ($\Theta$) directly from ground state charge density, the fundamental variable representing the spatial distribution of electrons in a system's lowest energy state. Determining elastic modulus from DFT requires a series of at least 15-20 calculations on strained unit cells. The Debye temperature requires a computationally intensive first-principles phonon dispersion calculation. Replacing those with a single charge density-based prediction has an obvious advantage, particularly for accelerated high-throughput screening of materials for specific applications.    

\begin{figure*}
\includegraphics[width=\linewidth]{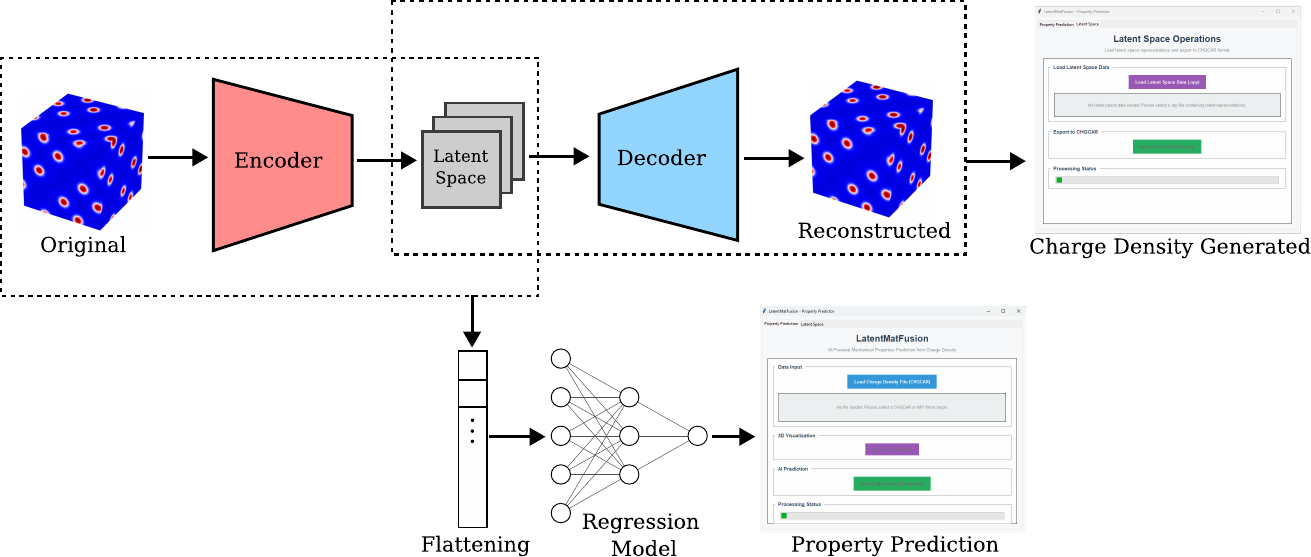}
\caption{The original charge density data (\(128 \times 128 \times 128\)) is passed through the encoder to obtain a compressed latent representation (\(16 \times 16 \times 16 \times 16\)). The autoencoder is trained by minimizing the reconstruction loss between the input and output. The compressed latent representation is utilized for two purposes: (a) One can decode and get back the original charge density, which is readable by the VESTA software~\cite{Mommako5060}, (b) It can be further flattened and used as input to a regression model for predicting materials properties. Two simple graphical user interfaces (GUIs) are provided for charge density extraction and materials property prediction. We have also provided a database of compressed latent charge densities (requiring only $\approx 1/117$ of the storage compared to the original charge densities), which can be downloaded and extracted by the GUI. The property prediction module can read the user-provided charge density for any new material not in the database and predict its properties. All the codes and data are available at \href{https://github.com/CMSLabIITK/LatentMatFusion}{GitHub}.}
\label{fig:overall_architecture}
\end{figure*}

The overall model architecture and its capabilities are illustrated in Figure~\ref{fig:overall_architecture}. To tackle the curse of dimensionality imposed by large $N\times N\times N$ data grids, which leads to substantial computational overhead and potential overfitting, our work integrates an autoencoder for efficient, unsupervised dimensionality reduction of the charge density. The autoencoder learns a compact, low-dimensional latent representation (a ``digital fingerprint'' of the electronic structure) that effectively preserves the essential, physically meaningful features while discarding redundant information. Once the original charge density data (\(128 \times 128 \times 128\)) goes through the encoder, it yields a compressed latent representation (\(16 \times 16 \times 16 \times 16\)). This allows us to provide a vast database of compressed latent charge densities (requiring only $\approx 1/117$ of the storage compared to the original charge densities), which can be decoded using a simple GUI to reconstruct the original charge density (\href{https://github.com/CMSLabIITK/LatentMatFusion}{GitHub}).

This compressed latent space then serves as the high-quality, reduced-feature input for the subsequent regression model. The pre-processing step not only renders the downstream regression task computationally more tractable but also, crucially, enhances the model's efficiency and predictive accuracy. We compare different regression models: the Attention-Enhanced CNN and LightGBM. We have also used composition-based material descriptors (Material Agnostic Platform for Informatics and Exploration, or MAGPIE~\cite {Ward_2016, D3DD00199G}) along with the latent charge density to enhance the model's predictive power. One can provide the charge density for any new material not in the database, and the trained model will predict its properties. A simple GUI is provided for materials property prediction (\href{https://github.com/CMSLabIITK/LatentMatFusion}{GitHub}). The following sections describe the technical details of the model.

\section{Input charge density calculation}

The charge density maps used as inputs for the deep learning models were generated from density functional theory (DFT) calculations performed using the Vienna Ab initio Simulation Package (VASP, version 5.4.4)\cite{Kresse1996}. The corresponding crystal structures were obtained from the Materials Project database~\cite{Jain2013}. The projector augmented wave (PAW) method within the generalized gradient approximation using the Perdew–Burke–Ernzerhof exchange-correlation functional (PAW–PBE)~\cite{PhysRevB.50.17953, PhysRevLett.77.3865} was employed to describe the electron–ion interactions. The electronic wavefunctions were expanded in a plane-wave basis set with an energy cutoff of 500 eV to ensure an accurate representation of the electronic charge density.

During structural relaxation, electronic self-consistency was achieved using a Gaussian smearing scheme to ensure stable convergence of metallic and semimetallic systems, with an energy convergence threshold of $10^{-5}$ eV. Structural relaxation was performed using the conjugate gradient algorithm, simultaneously optimizing ionic positions, cell shape, and volume until the residual forces were below $10^{-3}$ eV$\cdot$\AA$^{-1}$. Instead of using a fixed number of \textit{k}-points, reciprocal space sampling was controlled using a uniform \textit{k}-point spacing of 0.1~\AA$^{-1}$, ensuring consistent Brillouin-zone resolution for materials with different lattice parameters. Following structural convergence, the electronic charge density was recalculated using the tetrahedron method with Blochl corrections to obtain a high-fidelity, smearing-independent charge density. The resulting three-dimensional charge density distributions were extracted in volumetric form and used as inputs for subsequent regression analysis.

\begin{figure*}
    \centering
    \includegraphics[width=1\textwidth]{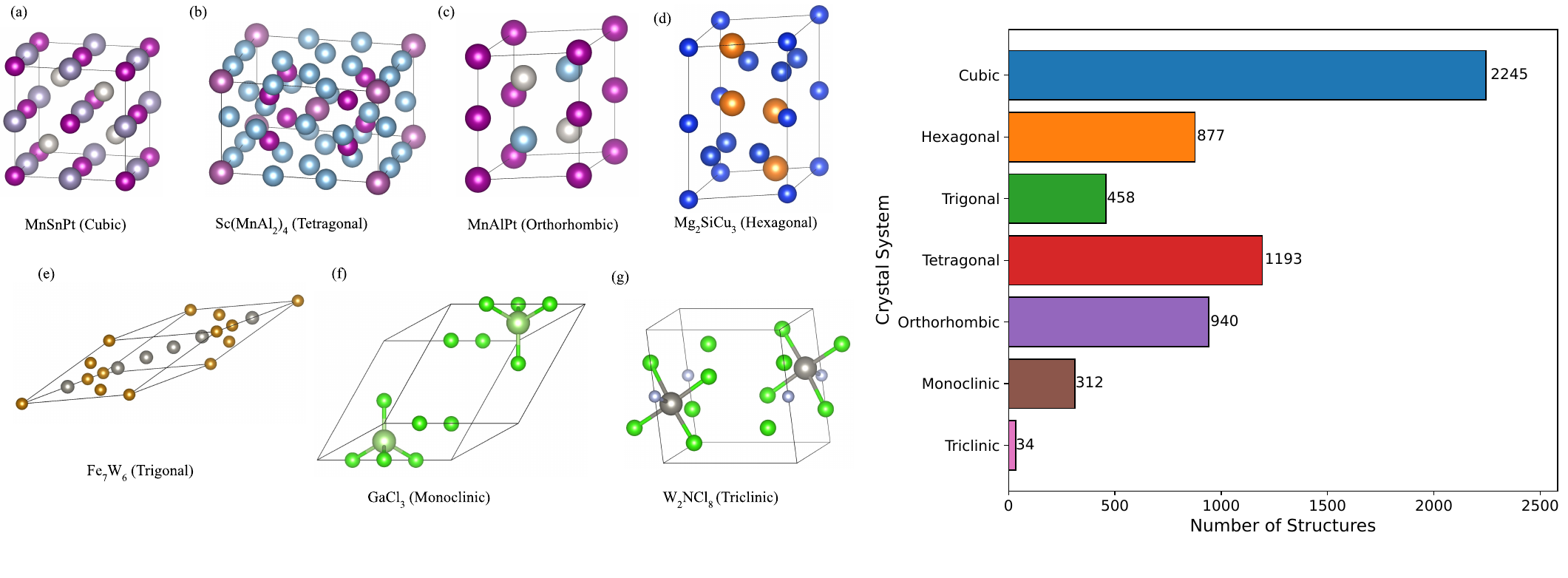}
    \caption{Representative crystal structures and their distribution across crystal systems. The panel on the left illustrates one representative structure from each crystal system, highlighting the diversity of structural motifs: (a) cubic, (b) tetragonal, (c) orthorhombic, (d) hexagonal, (e) trigonal, (f) monoclinic, (g) triclinic. The panel on the right presents a bar plot of the number of structures per crystal system, providing a quantitative overview of the dataset's composition.}
    \label{fig:crystals_fig}
\end{figure*}

The charge density values were transformed into array-based representations to enable efficient processing within the deep learning framework, using the Atomic Simulation Environment (ASE)~\cite{ase2002}. Owing to inherent structural variations among different materials (Figure~\ref{fig:crystals_fig}), the dimensions of these 3D arrays varied substantially across the dataset. To facilitate batch processing and convolutional operations, all input samples were padded to a uniform shape of \(128 \times 128 \times 128\). This dimension was chosen for its compatibility with the deep learning model architecture, as it enables two successive max-pooling operations with a pool size of 2 (specifically, \(16\times2^3=128\)), making it easier to apply the max-pooling operations efficiently. The input charge density data were not scaled during autoencoder training to preserve the original distribution. However, after obtaining the latent features from the trained encoder, they were standardized before training the regression model. This step was essential to ensuring stable, efficient learning. 

\section{Input property database}
\label{sec:Data Collection and Statistical Analysis}

\begin{figure*}
    \centering
    \includegraphics[width=1.0\textwidth]{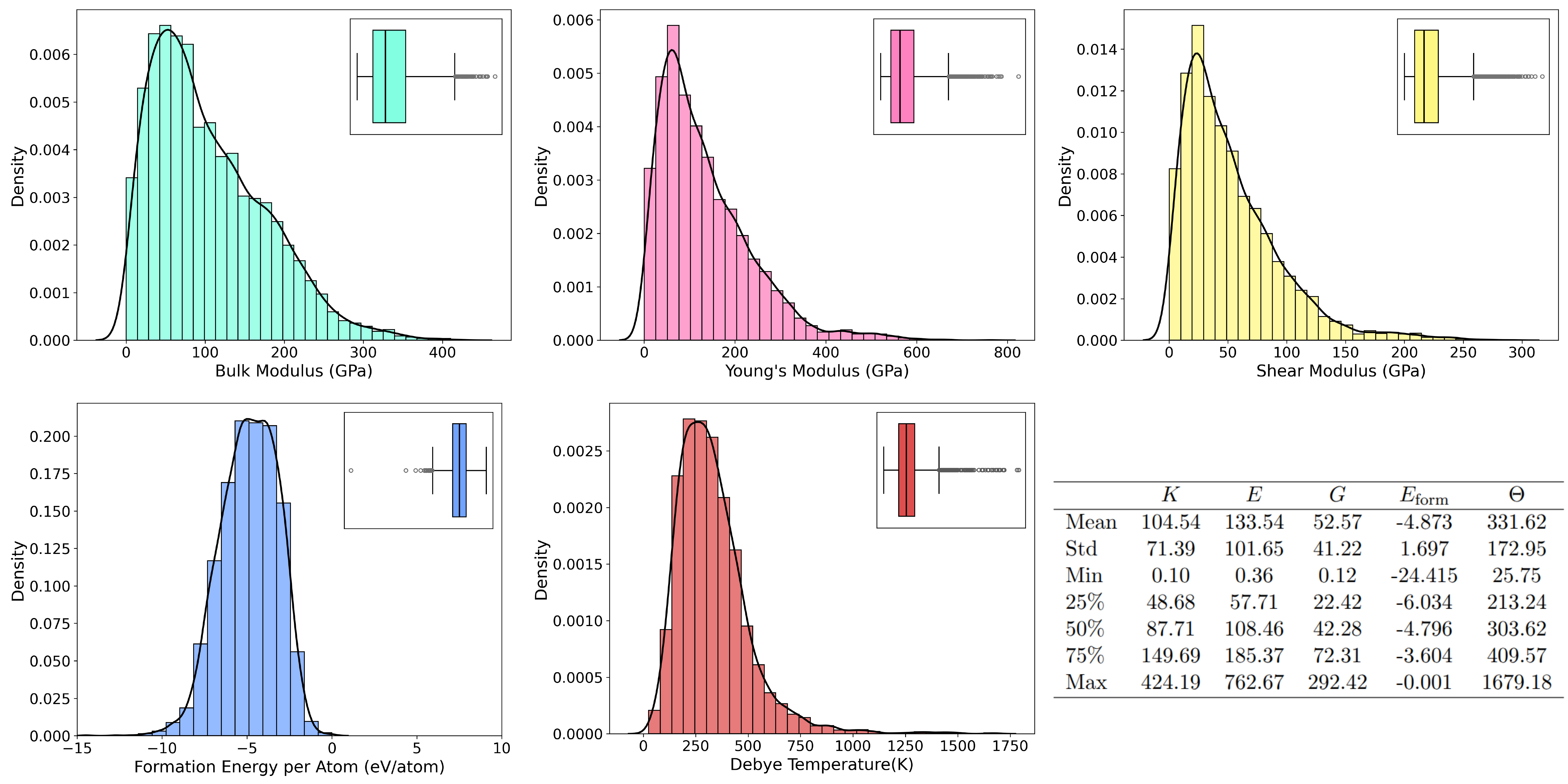}
    \caption{Statistical distributions of five key material properties: bulk modulus (K), Young's modulus (E), shear modulus (G), formation energy (E$_{\mathrm{form}}$), and Debye temperature ($\Theta$), are presented. The main panels display histograms of each property, overlaid with Kernel Density Estimation (KDE) curves (black solid lines). Inset box plots provide a complementary summary of the data, indicating the median, interquartile range, whiskers, and potential outliers. The bottom-right section summarizes key statistics of 6059 materials, including mean, standard deviation, percentiles, and extremes.}
\label{fig:hist_box}
\end{figure*}

The target properties considered in this work include the bulk modulus ($K$), shear modulus ($G$), Young’s modulus ($E$), formation energy per atom ($E_{\text{form}}$), and Debye temperature($\Theta$). Except for the formation energy per atom, the remaining quantities were sourced from the Materials Project database~\cite{Jain2013}, which provides results from standardized calculations for these mechanical and thermodynamic parameters. We calculated $E_{\text{form}}$ ourselves, using free atoms as a reference. The dataset encompasses a diverse range of crystal classes and space groups. The distribution of samples across different crystal classes is presented in Figure~\ref{fig:crystals_fig}. Their corresponding property distributions are visualized using histograms and boxplots in Figure~\ref{fig:hist_box}. Across all target properties, the dataset exhibits significant variability and strong non-Gaussian behavior. These plots highlight the chemically and mechanically diverse nature of the dataset, providing essential context for interpreting subsequent predictive analyses. The following texts present a detailed statistical analysis of the data.

The distribution of bulk modulus values spans a wide range, from approximately 0.1 GPa to 424.2 GPa, with a mean value of 104.5 GPa and a standard deviation of 71.4 GPa. The median bulk modulus is 87.7 GPa, indicating a moderately right-skewed distribution dominated by materials with low to intermediate stiffness. The interquartile range spans 48.7 GPa (25$^{th}$ percentile) to 149.7 GPa (75$^{th}$ percentile), highlighting the substantial diversity in volumetric stiffness across the dataset. The long tail toward higher bulk modulus values corresponds to a limited number of ultra-stiff materials, typically associated with strong covalent or mixed covalent-ionic bonding. The observed skewness possibly reflects the natural imbalance in available materials data, where mechanically soft and moderately stiff compounds are more prevalent than extremely incompressible systems.

Young’s modulus exhibits a broad range, from approximately 0.4 GPa to 762.7 GPa, with a mean of 133.5 GPa and a standard deviation of 101.7 GPa. The median value of 108.5 GPa lies noticeably below the mean, due to the right-skewed distribution. The interquartile range, spanning 57.7 GPa to 185.4 GPa, suggests that the majority of materials cluster within a moderate stiffness regime, while a small fraction of outliers exhibit exceptionally high tensile stiffness. 

The shear modulus distribution extends from values as low as 0.12 GPa to a maximum of 292.4 GPa, with a mean of 52.6 GPa and a standard deviation of 41.2 GPa. The median shear modulus is 42.3 GPa, indicating a concentration of materials with relatively low resistance to shear deformation. The interquartile range, spanning from 22.4 GPa to 72.3 GPa, reveals that shear stiffness is generally lower and more narrowly distributed than bulk and Young’s moduli.

The formation energy per atom exhibits a distinctly different distribution compared to the elastic properties, spanning from -24.415 eV/atom to values close to 0 eV/atom. The mean formation energy is -4.873 eV/atom, with a standard deviation of 1.697 eV/atom and a median value of -4.796 eV/atom. The narrow clustering of values around strongly negative energies indicates that thermodynamically stable compounds dominate the dataset. Such a distribution is characteristic of a curated materials database, where only experimentally reported or computationally stable compounds are included.

Debye temperature values span a wide range from approximately 25.8 K to 1679.2 K, with a mean of 331.6 K and a large standard deviation of 173.0 K. The median Debye temperature is 303.6 K, with an interquartile range extending from 213.2 K to 409.6 K. The outliers represent a smaller subset of compounds having exceptionally high Debye temperatures, typically associated with light elements and strong interatomic bonding.

\section{Latent representation of charge density}

\begin{figure*}
    \centering
    \includegraphics[width=\linewidth]{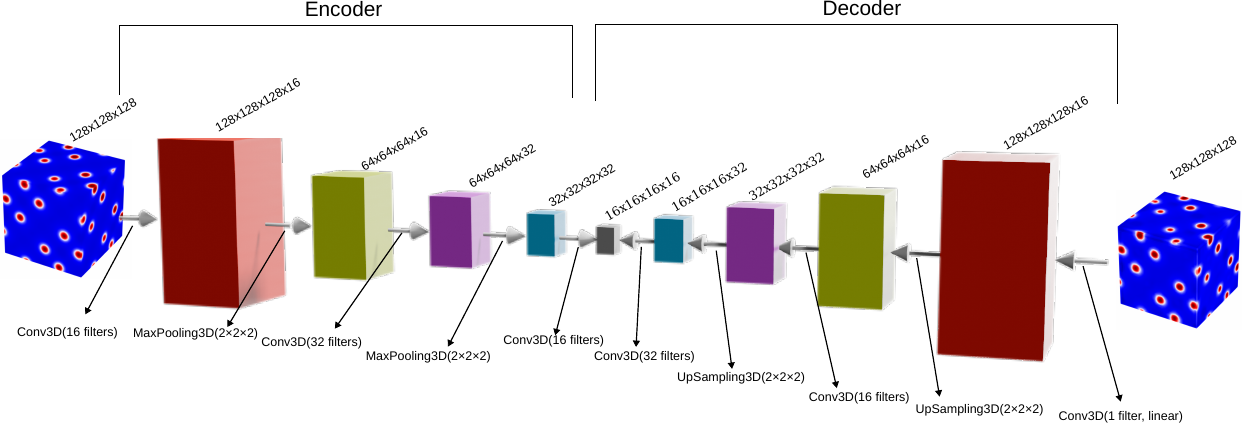}
    \caption{Schematic representation of the 3D convolutional autoencoder architecture. The encoder progressively reduces the input volume of size $128 \times 128 \times 128 \times 1$ through convolution and max-pooling operations, compressing it into a latent representation of size $16 \times 16 \times 16 \times 16$.}
    \label{fig:ae_schematic}
\end{figure*}

A 3D convolutional autoencoder was trained on raw charge density grids to obtain compact latent representations of the electronic structure. During autoencoder training, rotational data augmentation was applied to the charge density grids. Each sample was rotated along the principal axes to generate 12 distinct orientations, which were included in the training set. As shown in Figure~\ref{fig:overall_architecture}, the encoder compresses the high-dimensional input into a latent tensor, which is subsequently passed to a downstream regression model. This two-stage framework enabled accurate estimation of elastic properties, formation energy, and Debye temperature directly from charge-density data, without relying on hand-crafted descriptors.

The autoencoder is a fully convolutional neural network comprising two main parts: the encoder and the decoder. Both components utilize three-dimensional (3D) convolutional operations, making the architecture well-suited for handling the spatial complexity of volumetric charge density data. The architecture of the autoencoder is shown in Figure~\ref{fig:ae_schematic}. The encoder compresses the high-dimensional input volume of shape $(128 \times 128 \times 128 \times 1)$ into a compact latent representation of shape $(16 \times 16 \times 16 \times 16)$. This is achieved by applying two consecutive blocks, each consisting of a 3D convolutional layer followed by a ReLU activation and max pooling, and then one final 3D convolution with ReLU to produce the latent space. The decoder mirrors this structure, reconstructing the original input by applying two 3D convolutional blocks with ReLU activation and upsampling, followed by a final 3D convolution to recover the output volume. 

To ensure training stability and enhance generalization, each convolutional block is followed by batch normalization and dropout regularization. Batch normalization helps to accelerate convergence and mitigate internal covariate shift, while dropout randomly turns off a fraction of neurons during training, reducing the risk of overfitting. The final output of the decoder retains the same shape as the original input, enabling direct voxel-wise comparison during training. 

The dataset was randomly split into 80\% for training, 10\% for validation, and 10\% for testing to ensure robust model evaluation on unseen data. The training setup was carefully designed to strike a balance between learning efficiency and model generalization. These configurations aim to ensure stable convergence while preventing overfitting and optimizing training time. Early stopping based on the validation loss was used to avoid overfitting and optimize training time. The model's performance was continuously monitored on the validation set to ensure stable convergence and reliable generalization. The autoencoder was optimized by minimizing the mean squared error (MSE) loss, which computes the average squared difference between each voxel in the input and its reconstruction across the entire batch, depth, height, and width dimensions. Further details are provided in Figure S1, Supporting Information (SI).

\begin{figure} 
\centering 
\includegraphics[width =1.0\linewidth]{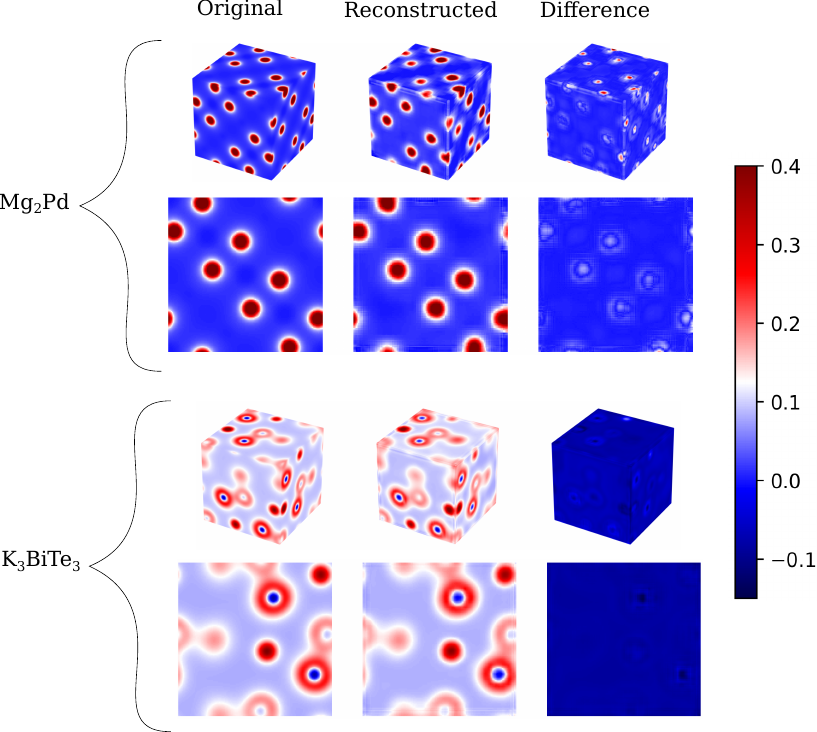}
\caption{Comparison of the charge density distributions for representative cubic compounds. The first column shows the original charge density, the second column presents the reconstructed charge density from the machine learning model, and the third column shows the difference between the original and reconstructed charge densities.}
\label{fig:charge_density_reconstructed}
\end{figure}

To evaluate the reconstruction performance of the trained autoencoder, two test compounds, $\mathrm{Mg_{2}Pd}$ and $\mathrm{K_{3}BiTe_{3}}$, were randomly selected, as shown in Figure~\ref{fig:charge_density_reconstructed}. The figure presents three-dimensional visualizations and two-dimensional cross-sectional slices of the original and reconstructed charge densities, along with their corresponding difference maps. The differences are consistently negligible across both samples, indicating that the autoencoder successfully learns a compact latent representation while preserving the essential structural features of the input data. This confirms that the latent space is sufficiently expressive for subsequent downstream tasks.

To evaluate model performance across the entire test set, reconstruction errors were computed. On average, the MSE was $2.1 \times 10^{-4}$, and the average MAE was $0.00402$. These values confirm that the autoencoder produces consistent, accurate reconstructions across diverse samples, with no significant deviations or outliers, thereby demonstrating the model’s ability to encode and regenerate complex volumetric features effectively.

Other than the selected $(16 \times 16 \times 16 \times 16)$ latent representation, two more latent configurations were explored. A strongly compressed model produced a latent tensor of $(8 \times 8 \times 8 \times 8)$, which reduced storage requirements but led to noticeable loss in the reconstructed charge densities [Figure~S2 and Figure~S3, SI], and it was discarded. Conversely, a higher-dimensional model with a latent space of $(32 \times 32 \times 32 \times 32)$ preserved fine-grained features and yielded slightly improved reconstruction quality [Figure~S2 and Figure~S3, SI]. When higher-dimensional latent representations were processed through appropriately adapted regression models for property prediction, this significantly increased computational cost without affecting the final outcome.  This exercise confirmed that the intermediate latent space dimension of $(16 \times 16 \times 16 \times 16)$ offers the optimal balance between reconstruction fidelity, predictive performance, and computational efficiency.

\section{Regression Model: LightGBM}

\begin{figure*}
    \centering
    \includegraphics[width=\linewidth]{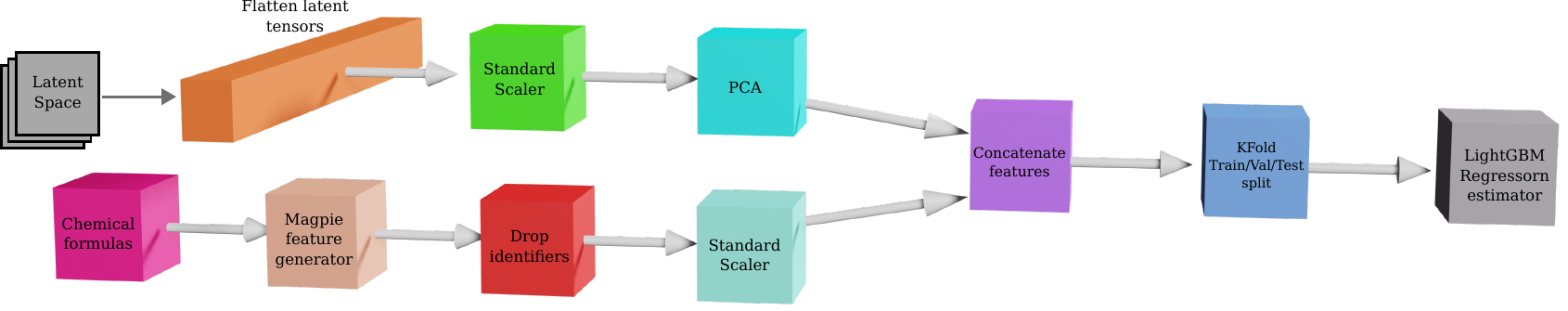}
    \caption{Schematic of the hybrid feature-fusion workflow for materials property prediction, combining PCA-reduced latent features from a deep-learning model with standardized MAGPIE compositional descriptors, followed by K-fold training and prediction using a LightGBM regressor.}
    \label{fig:schematic_prop_pred}
\end{figure*}

Property predictions were done using a hybrid tree-fusion learning framework that integrates latent charge density (dimension: $16\times 16\times 16 \times 16$, derived from DFT-calculated 3D charge density) with composition-based descriptors. This approach is designed to simultaneously capture local electronic-structure features and global chemical trends within a unified regression model. A schematic overview of the complete tree-fusion learning workflow, including feature generation, dimensionality reduction, feature concatenation, and model training, is shown in Figure~\ref{fig:schematic_prop_pred}.

As described in the previous section, latent representations of the charge density were obtained from a pretrained autoencoder and stored as three-dimensional tensors for each compound. These tensors were flattened into one-dimensional feature vectors and standardized using a zero-mean, unit-variance transformation. The dimensionality of the flattened latent space was further reduced using principal component analysis (PCA) to minimize feature redundancy, suppress noise, and preserve the dominant variance. A fixed number of 512 principal components was retained, capturing the majority of the explained variance and providing a compact yet information-rich representation of the latent electronic features.

In parallel, composition-derived MAGPIE descriptors were extracted for each compound, encompassing averaged elemental properties such as atomic mass, electronegativity, valence-electron count, and periodic table attributes [Section S1, SI]. These descriptors were independently standardized to ensure numerical compatibility with the PCA-reduced latent features. The final input feature vector was constructed by concatenating the PCA-compressed latent features with the scaled MAGPIE descriptors, yielding a fused representation that encodes both electronic charge density and compositional information. Prior to model training, numerical outliers in the target bulk modulus values were removed to improve model robustness and prevent undue influence from extreme values. The cleaned dataset was then partitioned using a shuffled five-fold cross-validation, yielding distinct training, validation, and test subsets. The validation set was used exclusively for hyperparameter optimization, while the test set remained completely unseen during model selection and evaluation.

The regression model was implemented using the Light Gradient Boosting Machine (LightGBM), a gradient-boosted decision tree ensemble particularly well-suited to high-dimensional, heterogeneous feature spaces. LightGBM iteratively constructs decision trees by minimizing a regularized least-squares loss function, enabling the efficient learning of complex non-linear interactions between latent and compositional features. Hyperparameters governing tree depth, number of leaves, learning rate, sub-sampling ratios, and regularization strengths were optimized using Bayesian optimization as implemented in the Optuna framework. The optimization objective was the validation root mean square error (RMSE), and early stopping was employed during training to mitigate overfitting.

After hyperparameter optimization, the final LightGBM model was retrained using the optimal parameter set with a larger number of boosting iterations. Model performance was assessed on the independent test set using the coefficient of determination ($R^2$), mean absolute error (MAE), and root mean square error (RMSE) [Section S2, SI]. All computations were performed with fixed random seeds to ensure full reproducibility. The trained model was serialized for subsequent deployment and analysis, enabling consistent property-value predictions for previously unseen compounds.

\section{Regression Model: Attention based 3D CNN}
\begin{figure*}
    \centering
    \includegraphics[width=1.1\linewidth]{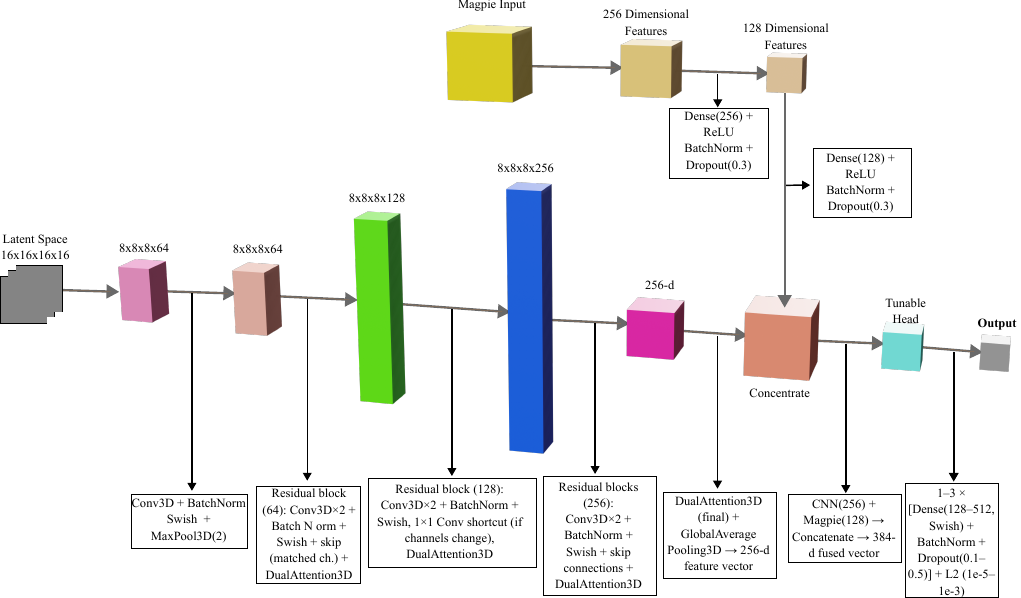}
    \caption{Attention 3D CNN model: The architecture consists of stacked residual 3D convolutional blocks with batch normalization and Swish activation, where dual-attention modules selectively emphasize physically relevant spatial regions and feature channels within the charge-density representation. Global average pooling condenses the learned descriptors into a fixed-length feature vector, which is combined with Magpie descriptors to incorporate chemical information. The resulting hybrid representation is processed through fully connected layers with batch normalization and dropout to produce the final output.}    
    \label{fig:attn_cnn_schematic}
\end{figure*}
The materials were represented using volumetric electronic structure descriptors in conjunction with composition-based MAGPIE features [Figure~\ref{fig:attn_cnn_schematic}]. The volumetric latent tensors were normalized to ensure numerical stability, while the composition-derived descriptors were standardized using zero-mean, unit-variance scaling. Target values were scaled to the unit interval during training and subsequently transformed back to their original physical units for evaluation.

The 3D CNN model is composed of stacked residual convolutional blocks with batch normalization and Swish activation. Dual-attention modules were incorporated within the network to enhance the representation of physically relevant spatial regions and feature channels in the volumetric data. Global average pooling was applied to obtain a fixed-length descriptor of the electronic charge density. In parallel, composition-derived MAGPIE descriptors were transformed using a multilayer perceptron with fully connected layers, batch normalization, and dropout regularization. The outputs of the volumetric and compositional branches were concatenated to form a hybrid representation, which was further processed through a regression head composed of one to three dense layers before producing the final target property prediction.

Model hyperparameters, including the depth and width of the regression head, regularization strength, dropout rates, optimizer choice, and learning rate, were optimized using random search with validation loss as the objective. Training was performed using the mean squared error loss function with early stopping to prevent overfitting. The dataset was divided into training, validation, and test subsets using fixed random seeds to ensure reproducibility. The final model's performance was assessed on the independent test set using $R^2$ and MAE.

\section{Results and Discussions}
\label{sec:discussions}
\begin{figure*}[ht]
    \centering
    \includegraphics[width=\textwidth]{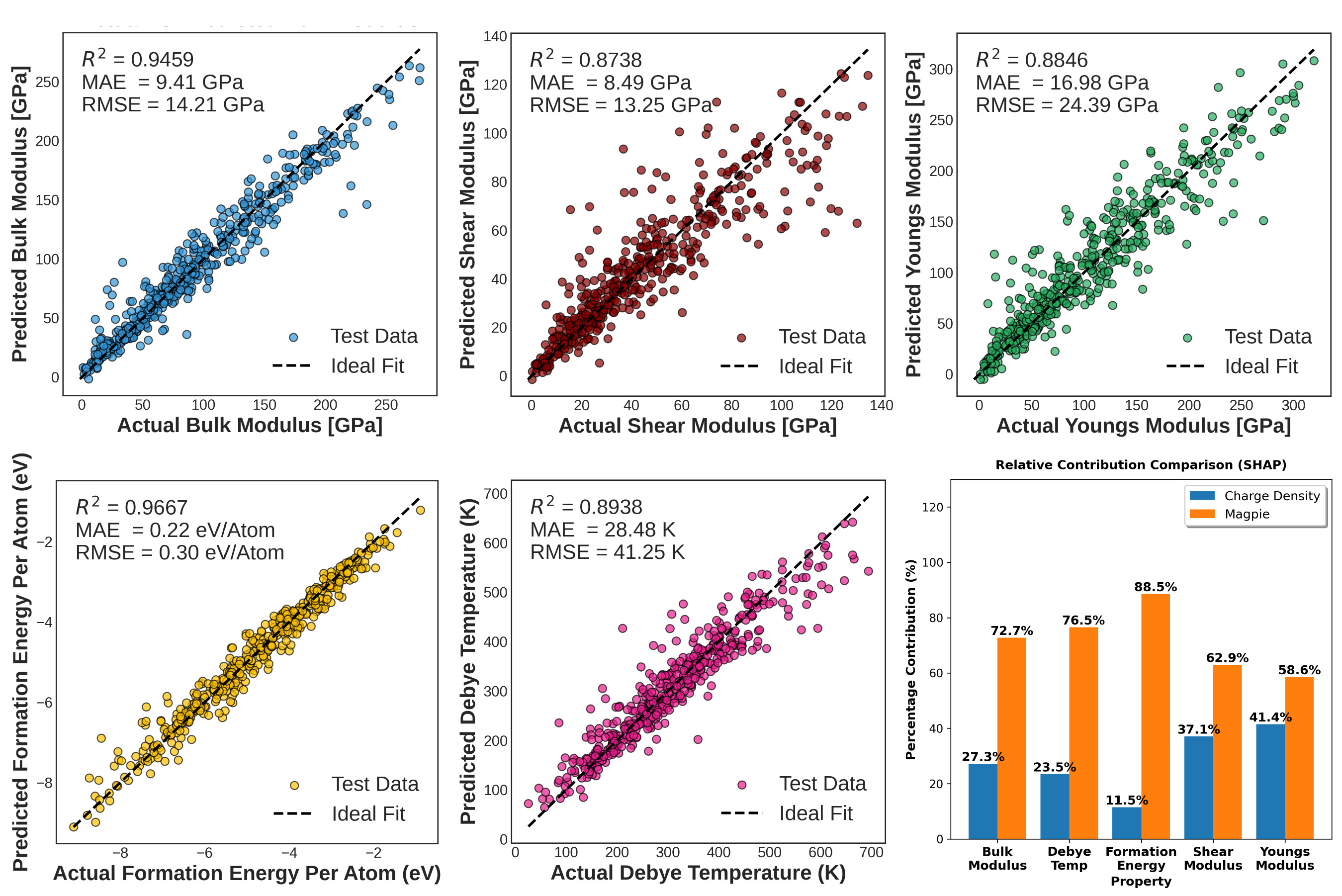}
    \caption{LightGBM model: Actual versus predicted values for the five material properties considered: bulk modulus, Young’s modulus, shear modulus, formation energy, and Debye temperature. The diagonal black line represents the ideal case of perfect prediction. The bottom-right panel compares the impact of charge density and MAGPIE features on predictions, indicating that the latter contribute more across the evaluated properties.}
    \label{fig:actual_predicted}
\end{figure*}

The effectiveness of the first hybrid ML architecture (latent charge density + MAGPIE features, LightGBM regression) is quantitatively evaluated using the actual-versus-predicted plots shown in Figure~\ref{fig:actual_predicted}, which provide a direct assessment of model accuracy across the entire test dataset. Evidently, the model shows excellent performance across all five target properties. Among the elastic properties, the bulk modulus shows the highest accuracy, with an $R^2$ of 0.94. Young’s modulus and shear modulus are also predicted with high accuracy, achieving $R^2$ values of 0.88 and 0.87, respectively. The formation energy per atom exhibits the strongest correlation between predicted and actual values, with an $R^2$ of 0.96. The Debye temperature, which reflects lattice vibrational characteristics and elastic stiffness, is also predicted with reasonable accuracy ($R^2 = 0.89$). The bottom-right panel of Figure~\ref{fig:actual_predicted} compares the impact of charge density and MAGPIE features on predictions, estimated using SHAP (SHapley Additive exPlanations), indicating that the latter contribute more across the evaluated properties, although they still complement each other. 

\begin{figure*}[ht]
    \centering
    \includegraphics[width=\textwidth]{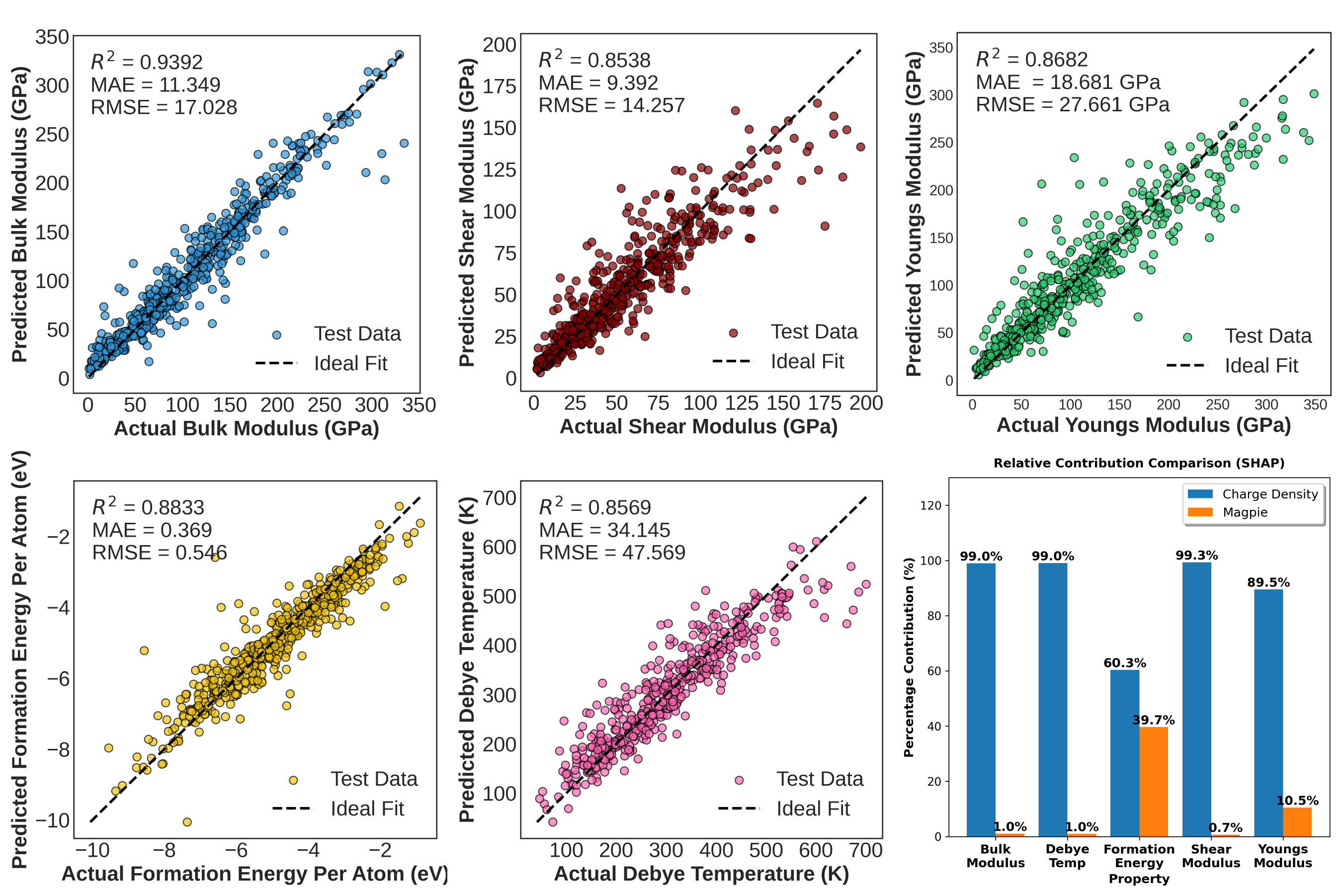}
    \caption{Attention CNN model: Actual versus predicted values for the five material properties considered: bulk modulus, Young’s modulus, shear modulus, formation energy, and Debye temperature. The diagonal black line represents the ideal case of perfect prediction. The bottom-right panel compares the impact of charge density and MAGPIE features on predictions, indicating that the former contributes more across the evaluated properties.}
    \label{fig:actual_predicted1}
\end{figure*}

The effectiveness of the second hybrid ML architecture (latent charge density + MAGPIE features, Attention CNN regression) is quantitatively evaluated using the actual-versus-predicted plots shown in Figure~\ref{fig:actual_predicted1}, providing a direct assessment of model accuracy across the entire test dataset. Similar to the previous model, this also shows very good performance across all five target properties. Among the elastic properties, the bulk modulus shows the highest accuracy, with an $R^2$ of 0.94. Young’s modulus and shear modulus are also predicted with high accuracy, achieving $R^2$ values of 0.87 and 0.85, respectively. The formation energy per atom achieves an $R^2$ of 0.88, while the Debye temperature is also predicted with reasonable accuracy ($R^2 = 0.86$). The bottom-right panel of Figure~\ref{fig:actual_predicted1} compares the impact of charge density and MAGPIE features on predictions. Unlike the LightGBM regression model, the charge density features contribute more across the evaluated properties in the case of the Attention CNN regression model.

\begin{table}
    \centering
    \begin{tabular}{|c|c|c|c|c|c|}
    \hline
    Input & K & E & G & E$_{form}$ & $\Theta$ \\
    \hline
    Latent CD + CNN & 0.90 & 0.79 & 0.80 & 0.68 & 0.73\\
    \hline
    Magpie + Att CNN & 0.92 & 0.85 & 0.85 & 0.88 & 0.90\\
    \hline
    Latent CD + MAGPIE + LightGBM & 0.94 & 0.88 & 0.87 & 0.96 & 0.89\\
    \hline
    Latent CD + MAGPIE + Att CNN & 0.94 & 0.87 & 0.85 & 0.88 & 0.86\\ 
    \hline
    \end{tabular}
    \caption{$R^2$ values obtained from different ML models. The first two rows highlight the models trained only on the Latent charge density (CD) and only on Magpie features. The last two rows highlight the models trained with hybrid features (charge density + MAGPIE), and predictions are done using LightGBM and Attention CNN. Our results compare well with Crystal Graph Convolutional Neural Networks (CGCNN)~\cite{CGCNN2018}, which achieves $R^2$ values of 0.93, 0.89, 0.88, and 0.94 for K, E, G, and $\Theta$, respectively.}
    \label{tab:R2value}
\end{table}

Having observed contrasting behavior of the two hybrid models in terms of feature importance analysis, we now analyze further to quantify the effect of charge density and Magpie descriptors systematically. For this purpose, two additional deep learning models were trained and evaluated on the same dataset using the same train-validation-test splits. The first model [Section S3, Figure S4, SI] is trained only on the Latent charge density (CD). As reported in Table~\ref{tab:R2value}, the model performs reasonably well for elastic moduli prediction, but not so well for formation energy and Debye temperature. The underlying bonding characteristic is encoded in the charge density, allowing a reasonable prediction of elastic moduli. Debye temperature, on the other hand, not only depends on elastic moduli, but also is inversely proportional to the molecular weight, which is not encoded in the Latent CD. Similarly, formation energy depends on factors such as electronegativity, ionization energy, and electron affinity, which are not encoded in the Latent CD. As shown below, inclusion of MAGPIE features significantly improves the prediction of Debye temperature and formation energy. 

The second model [Section S4, Figure S5, SI] is trained only on the MAGPIE features and outperforms the Latent CD-only model comprehensively. However, compared to end-to-end models based solely on composition, the charge density provides an explicit electronic representation of bonding, charge redistribution, and screening effects. This representation improves physical interpretability and enables potential transferability in certain cases where composition-only models may struggle. For example, in the case of polymorphs, the chemical formula remains the same, although the charge density changes. In such a case, charge-density-based predictions are the only option. This further underscores the usefulness of a hybrid model combining CD and MAGPIE features. Overall, incorporating charge density–based features significantly enhances the model’s ability to capture the underlying physics governing material properties.

To verify the model robustness, we have carried out multiple tests. First, we confirmed the rotational invariance by rotating the input charge density by different angles and predicting properties using the rotated charge density. The predicted values lie within a narrow range, with a standard deviation up to $\approx 5\%$ [see Fig~S6, SI], indicating that our model is robust to rotations. Next, we also confirmed that the outcome does not depend on the choice of the unit cell. For Cu$_3$Al and Mg$_2$Ca, we have compared the results of a small simulation cell and a big supercell, and obtained a very good agreement [see Fig~S7, SI].

Let us highlight the significant difference between our method and similar approaches used to predict elastic properties from charge densities of materials~\citep{mirzaee2024elastic, Zhao2020} [see Figure~S8, SI]. Other previously reported combined DFT-ML work dealt with a specific type of material, like face-centered cubic high entropy alloys by Mirzaee \emph{et al.}~\cite{mirzaee2024elastic} or face-centered cubic materials by Zhao \emph{et al.}~\cite{Zhao2020}, and lacks the generality of the present work in terms of type of bonding and crystal symmetry.  Similarly, Sidnov \emph{et al.} proposed a two-stage CatBoost-based framework for predicting elastic properties of BCC Ti and Zr alloys \cite{sidnov2024machine}, and Kim \emph{et al.} employed Gradient-Boosted Trees (GB-Trees) to predict the bulk and shear moduli of the high-entropy alloy Al$_{0.3}$CoCrFeNi~\cite{KIM2019124}. The dataset's restricted nature limits its applicability to more general crystal systems.

Notably, many descriptor-based studies consistently identify valence electron configuration as a dominant feature governing elastic behavior \cite{VAZQUEZ2022117924, REVI2021110671, Grant2022, Khakurel2021, MEI2023112249}, highlighting the central role of electronic interactions in determining mechanical properties. These interactions are inherently encoded in the electronic charge density distribution, motivating the development of direct charge-density-based learning approaches. Ground state electron charge density $\rho(r)$ is the central, fundamental quantity in DFT, as the Hohenberg-Kohn theorems state that all ground-state properties, including energy, forces, and electronic structure, are unique functionals of $\rho(r)$. Thus, DFT-based charge density is a natural choice, as it is guaranteed to provide a rich set of features for property prediction.

To facilitate the practical adoption of the proposed framework, a Python-based GUI (\href{https://github.com/CMSLabIITK/LatentMatFusion}{GitHub}) has been developed that enables direct interaction with the trained models. The GUI serves two primary purposes. First, it allows users to load three-dimensional electronic charge density data directly from VASP \texttt{CHGCAR} files and predict key material properties, including bulk modulus, Young’s modulus, shear modulus, formation energy per atom, and Debye temperature, using either the trained LightGBM model or the Attention CNN model. This functionality eliminates the need for manual data preprocessing and provides an efficient pathway for rapid property evaluation from first-principles electronic structure calculations. A model VASP \texttt{INCAR} file has been provided in Section~S5, SI, showing the recommended parameters, particularly \texttt{ENCUT} and \texttt{KSPACING}.

The second utility of the GUI enables users to load compressed latent-space representations of 6059 materials generated by the autoencoder and reconstruct the corresponding three-dimensional charge density fields. This might be a valuable ground-state charge density dataset readily available to researchers and easily accessible due to its compressed format, which would otherwise require massive storage. Together, these functionalities demonstrate the practical utility of the proposed deep learning framework and highlight its potential for integration into high-throughput computational materials workflows.

In conventional DFT workflows, the computational cost of obtaining elastic properties and related quantities depends on the symmetry. For example, in cubic systems, three independent deformation modes are required to determine the elastic stiffness matrix. Each deformation mode typically involves at least seven total-energy calculations at different strain magnitudes, resulting in approximately 21 calculations. For lower-symmetry structures, computational effort increases substantially; for example, 21 independent deformation modes are required in triclinic crystals, with at least 7 calculations per deformation, resulting in approximately 147 total DFT calculations. The method described in this work can predict multiple properties from a single self-consistent DFT calculation, requiring $\approx$ 1/25 (or even less for low-symmetry systems) of the computational resources required for full-fledged DFT calculations.

\section{Conclusions}
\label{sec:conclusions}
This work establishes a robust and scalable deep learning framework that directly links three-dimensional electronic charge density from \textit{ab initio} calculations to macroscopic mechanical and thermodynamic properties. By combining a 3D convolutional autoencoder for unsupervised dimensionality reduction with a regression model, the proposed approach effectively bridges the quantum-mechanical electronic structure and material response. 

The 3D convolutional autoencoder compresses high-dimensional charge density grids ($128\times128\times128$) into a compact latent representation ($16\times16\times16\times16$) while preserving physically meaningful features, as confirmed by negligible reconstruction errors for representative compounds.

Composition-based MAGPIE descriptors are combined with the latent charge-density, and the combined feature set is used for the regression. The resulting hybrid model consistently outperforms solely charge density-based deep learning approaches, achieving excellent predictive accuracy across all target properties, using LightVBM and Attention CNN-based regression. 

The ability to predict multiple properties from a single self-consistent DFT output, together with the developed Python-based graphical user interface (GUI), highlights the potential of this approach for high-throughput and practical materials screening, requiring $\approx$ 1/25 the computational resources of full-fledged DFT calculations. Overall, this study demonstrates that electronic charge density is a unified, transferable, and physics-informed descriptor that can capture complex structure-property relationships within a single learning framework.

\section{Data and Software Availability}
Data and code are available on GitHub \href{https://github.com/CMSLabIITK/LatentMatFusion}{https://github.com/CMSLabIITK/LatentMatFusion}.

\section{Acknowledgement}
We acknowledge the National Super Computing Mission (NSM) for providing computing resources of ``PARAM Sanganak'' at IIT Kanpur, which is implemented by CDAC and supported by the Ministry of Electronics and Information Technology (MeitY) and the Department of Science and Technology (DST), Government of India. We also thank the ICME National Hub, IIT Kanpur, and CC, IIT Kanpur, for providing an HPC facility.

\section{Supporting Information}
Additional details are provided regarding: Training and validation loss curve of the autoencoder model, Charge density reconstruction with different latent dimensions, Details of MAGPIE features, Details of loss function, Charge density-only CNN model, MAGPIE-only ML model, Model robustness in terms of rotation and cell size, Comparison with literature, Model VASP \texttt{INCAR} file.  

\bibliography{reference}

\end{document}